\documentstyle [12pt]{article}
\title{
Ferromagnetism below the Stoner limit in La-doped SrB$_6$.}
\author{T. Jarlborg\\
DPMC, University of Geneva, \\
24 Quai Ernest-Ansermet, CH-1211 Gen\`eve, Switzerland\\}
\begin{document}
\baselineskip 5mm
\maketitle


\begin{abstract}
Spin-polarized band calculations for LaSr$_7$B$_{48}$ show a weak ferro-magnetic state.
This is despite a low density-of-states
(DOS) and a low Stoner factor. The reason for the magnetic state is found
to be associated with a gain in potential energy in addition to the
exchange energy, as a spin-splitting is imposed. An impurity like La DOS
is essential for this effect. It makes a correction to
the Stoner factor, and provides an explanation of the recently
observed weak ferro-magnetism in doped hexaborides.

\end{abstract}


Recent experimental works on doped hexaborides 
show many surprises \cite{ott1,deg}. Among the unexpected properties is
the observation of a very weak ferromagnetic state that persists up to large
temperatures when small amounts of Ca are replaced by La in the hexaboride
system, La$_x$Ca$_{(1-x)}$B$_6$ \cite{you}. This is surprising since the
density-of-states at the Fermi energy, $N(E_F)$, and the Stoner
factor are expected to be small in this system. 
Band calculations \cite{mass} in the  local density approximation (LDA) \cite{lda}
on SrB$_6$ and CaB$_6$ confirm that the DOS is low, with $E_F$ falling in a
valley of the DOS, and spin-polarized calculations for EuB$_6$ show a very
weak spill-over of the spin-density from the Eu-f spin to the rest
of the valence band.
Electronic structure calculations based on the local (spin) density
approximation, either by spin-polarized calculations or by
calculations of the Stoner factor $\bar{S}$, are in general able to detect if a
system is ferromagnetic or not \cite{hunt,jan,c15}. Because the Stoner factor for La$_x$Sr$_{(1-x)}$B$_6$
is far below the limit for magnetism,  it is tempting to suggest that the
observed magnetism is not a band magnetism, but something else, perhaps excitonic \cite{ric}
or a polarization of the dilute electron-gas \cite{you,ort}.
However, for Ce
it is found that the Stoner factor underestimates the tendency towards magnetism,
because the stabilization energy is not only coming from exchange, and the Stoner
criterion has to be corrected for a potential energy \cite{tj98}. Even if this
correction is small in Ce, there is nothing in principle that prevents a large correction
to the Stoner model in other systems. In this letter we show that La doped SrB$_6$ is such
a system. According to the traditional understanding of ferromagnetism, the exchange energy 
is the source of a magnetic transition,
but here we find that the band structure is such that the exchange splitting triggers off
charge transfers that contribute to the total gain in energy via the Coulomb interaction.

Electron energies are calculated 
by the linear muffin-tin orbital (LMTO) method using the LDA.
The structure of SrB$_6$ is simple cubic, with one Sr in the corner, and a B$_6$ octahedron
in the center. The internal parameter $z$ which determines the shortest distance from one of the B-atoms
to a cube surface, is in these calculations 0.215 of the lattice constant. Additional 'empty' spheres
are inserted at the cube edges, between the Sr-atoms, in order to account for non-sphericity
of the potential in the open part of the structure. 
Calculations where made for the basic unit cell, a double cell 
and a large supercell of 8 cubes (2 cube lengths in x,y and z-direction), containing totally
56 atoms and 24 empty spheres.
The basis set for the largest
unitcell includes s,p,d for the atoms, and s,p for empty spheres. The small cells include also
f-states for the Sr-sites. The occupied f-part is always very small, even for the La impurity site.  
The lattice constant is fixed at 7.93 a.u. in each calculation. No lattice relaxation around the La impurity
is considered. The number of k-points varies from
286 to 20 irreducible k-points for the unit cells of 10 and 80 sites, respectively.

The bandstructure for undoped SrB$_6$ agree well with the one in ref. \cite{mass}.
The paramagnetic (non-spinpolarized) DOS from the supercell calculation for LaSr$_7$B$_{48}$ is shown in
fig. \ref{fig:dos}. The total DOS at $E_F, N(E_F)$, about 48 states/cell/Ry, is very concentrated on
the single La site, with $N_{La}(E_F) \sim$ 19 states/cell/Ry, cf Table 1. On the remaining 55 sites, each Sr has
on the average 1.0 and each B roughly 0.4 states/cell/Ry, respectively.
By comparison, a rigid-band model obtained by adding one electron to the band structure without the La
impurity (cf. fig. \ref{fig:dos}) gives very different values: Of the total DOS, 32 states/cell/Ry,
each Sr has 1.4 and each B 0.25 states/Ry.   
Clearly, La behaves a localized impurity site with a
very large local $N_{La}(E_F)$, much larger than on Sr. The additional valence electron remains
located on the La, as can be seen in Table 2. 
By comparing the charges in the undoped and doped supercell one finds that
the remaining Sr$_7B_{48}$-spheres receive no additional charge. The difference in charge between the two systems
is never larger than 0.01 electron per site, among these sites. The derivative of the DOS becomes very large, 
about 1400 states/cell/Ry$^2$,
out of which 700 is coming from the La site, compared to about 30 for each Sr site when the rigid-band model is
applied to the undoped bandstructure. This asymmetric and uneven distribution of the DOS near $E_F$ 
is a condition for real space charge transfers as T increases and as an exchange splitting
is imposed. The transfer goes in both cases to the La site from the rest of the system.

Spin-polarized calculations for pure SrB$_6$, and for a double cell SrLaB$_{12}$,
do not give stable moments of a significant size. The supercell calculation of LaSr$_7$B$_{48}$ using 20 k-points
finds a small moment of 0.10 $\mu_B$/cell at low temperature. 
The ferromagnetic state is about 10 mRy/supercell lower in total energy than the paramagnetic solution. 
If fewer number of k-points are used
one finds that the (paramagnetic) DOS and its energy derivative near $E_F$ are larger, cf Table 1. The spin
moment increases to 0.24 $\mu_B$/cell for the lowest number of k-points, 
with a total energy 11 mRy lower than the paramagnetic state.
The moment is always largest on La, with more than 50 percent of the
total moment on and near the La impurity. The moment on B-sites
far from the impurity is negligible. This fact and the uneven distribution of local DOS show that
a rigid-band picture of dilute La-doped hexaborides does not apply. The bands near $E_F$
are modified by the spacially non-uniform perturbation of the potential around the impurity site.

The original Stoner model \cite{hunt} can be applied together with ab-initio paramagnetic 
 band results in order to study the spin susceptibility or
the onset of magnetism  \cite{jan,c15}. 
 An applied exchange splitting, $\xi$, leads to a transfer from minority spins
to majority spins, and a loss 
in kinetic energy $\Delta K = N \xi^2$, where $N$ the DOS
at $E_F$ is assumed to be constant. The same spin transfer leads to a gain in
exchange energy $\Delta E_x = N^2 I_s \xi^2$, where
 the exchange integral, $I_s$, can be calculated within LDA \cite{jan}.
 If $\Delta E_x \geq \Delta K$, there will be a total gain in energy and a
ferromagnetic transition can occur.  
By elimination of factors of $N$ and $\xi^2$
one obtains the Stoner criterion, 
$\bar{S}=NI_s \geq 1$.  
If a partial DOS is not constant near $E_F$, there will be a possibility of charge transfers as function
of $\xi$, and additional contributions to the total energy. As in ref. \cite{tj98}, it is here possible to
identify one sub-band, the La-impurity (mainly d) band, which has a large derivative of its DOS, $N'_{La}(E_F)$.
The total DOS at $E_F$ is $N=N_{La} + N_v$, where $N_v$ is the rest of the DOS. 
 By neglect of all other derivatives one can make a
model of the charge transfers as function of $\xi$ and $k_BT$. From ref \cite{tj98},
we get the charge transfers
$\Delta q_{\xi}=
 N_vN'_{La} \xi^2/N$ and $\Delta Q=N_vN'_{La} \pi^2 (k_BT)^2/6N$, 
respectively, where it is assumed that $N'_{La} \cdot k_BT$ and $N'_{La} \cdot \xi$ are small. 

Next we calculate the Coulomb energy associated with the charge transfer, $\Delta E(T) = U_0 \Delta Q$, from
self-consistent paramagnetic calculations at different T.
The charge density is  $\rho(T,r) = 
 \sum_{{\vec k},j} f(\epsilon_{\vec k}^j,E_F,T) \mid \Psi(\epsilon_{\vec k}^j,r) \mid^2$ , where $f$ is 
 the Fermi-Dirac function,
and $\Psi (\epsilon_k^j,r)$ is the wavefunction for point $k$ and band $j$.
  The electronic total energy, $E_{tot}$, contains 
kinetic, Coulomb and exchange-correlation terms:
\begin{equation}
E_{tot}(T)= \int f \epsilon N(\epsilon) d\epsilon -  
\int (\frac{1}{2} V_e + \varepsilon - \mu) \rho d^3r + E_M
\label{eq:etot}
\end{equation}
$V_e$ is the electronic part of the Coulomb potential, $E_M$ is the Madelung part of the electrostatic
energy, and
 $\varepsilon$ and $\mu$ are the exchange-correlation energy and potential, respectively.

The first
term in Eq.~\ref{eq:etot}, the kinetic term, becomes
$E_K=\frac{\pi^2}{3}(k_B T)^2 N(E_F)$, giving the usual Sommerfeld term in the specific
heat $C_{el}=\frac{\partial E_K}{\partial T}=\gamma T$. 
The difference, $E(T)$,
between the fully calculated $E_{tot}$ in eq.  \ref{eq:etot} and $E_K$ defines 
of the non-kinetic contribution due to charge transfers:
\begin{equation} 
E(T)=\int f \epsilon N(\epsilon) d\epsilon -E_{tot}(T) = \int (\frac{1}{2} V_e + \varepsilon - \mu) \rho d^3r 
- E_M(T)
\label{eq:detot}
\end{equation}
and
\begin{equation} 
\Delta E(T) = E(T)-E(0)
\label{eq:delta}
\end{equation}

Calculations of $E(T)$ need to be selfconsistent because of relaxation, while $E_K$ and the Stoner product
are only moderately affected by non-linear response effects from the system.
At low T is $\Delta E$ nearly proportional to $\Delta Q$,  $\Delta E \approx U_0 \Delta Q$,
while it saturates for higher T.
The calculated parameter $U_0$ varies between 1 and 2.5
Ry/el./La for $k_BT$ in the interval 0.5 to 15 mRy. This is from calculations using 10 irreducible k-points.
It is probable that the main part of $\Delta E$ is
coming from the Madelung term, because of the sizeable charge transfer to La. This value of $U_0$ is
calculated witout taking into account lattice expansion, but the mechanism of large charge transfers
is expected to have an influence on thermal expansion and specific heat.

The charge transfer towards La is a result of the increasing La-DOS
 near $E_F$ (fig. \ref{fig:dos}).
A very similar charge transfer pattern  will result from
 a spin splitting, $\xi$, of the para-magnetic bands.
The similarity between the two types of charge transfer permits us to
 write $\Delta E_{\xi} = U_0 \Delta q_{\xi}= U_0 N'_{La} N_v \xi^2/N$. With this energy 
included, one obtains a modified
Stoner criterion \cite{tj98}:

\begin{equation}
\label{eq:msto} 
\bar{S}_U = N I_s + U_0 N'_{La} N_v /N^2  \geq 1    
\end{equation}

 The calculated Stoner product $N I_s$ is 0.21, $N'_{La} \sim$ 730 per Ry$^2 \cdot$ cell, 
 N $\sim 48$ and $N_v \sim 29$ (per Ry $\cdot$ cell), 
to give
$\Delta q_{\xi} \approx 440 \xi^2$  and $\Delta Q = 720 (k_BT)^2$,  with $\xi$ and $k_BT$ in Ry.
The values make the Coulomb term in eq. \ref{eq:msto} very large, about 14. However, even if the Coulomb
energy represents a major correction in this case, there are several
facts indicating that other energies are involved as well. 
First, the exchange splitting $\xi$ is not uniform over all sites. The spin-polarized calculation 
shows that $\xi$ is finite only at or very near to La, and the energy of a
variable $\xi$ is not counted. Second, although the amount of charge transfers
due to $\xi$ and $k_BT$  are similar, there are long tails of 
the Fermi-Dirac distribution which make the radial dependencies of $\Delta q_{\xi}$ and $\Delta Q(T)$ different.
Third,
 the effective derivative of the La-DOS, $N'_{La}(\epsilon,T)$, (taken as $\int (-\frac{df}{d\epsilon})
 N'_f(\epsilon) d\epsilon$), is likely to be
largest at low T, because the bare DOS has not the same large derivative over a wide energy range. 
 The mechanism of charge transfer and magnetism is therefore expected to disappear at some large T. 
This is confirmed in the spinpolarized results. By raising $k_BT$ to 3 mRy in the calculation
with 10 k-points, the magnetic moment goes
to zero. It is also interesting to note that the value of $N'_{La}$ is larger in the calculation
with fewer number of k-points, as shown in Table 1. 
The number of k-points are insufficient for a good convergence in this case,
but the variations of $N'_{La}$ and the magnetic moment in the spin-polarized results
as a function of the number of k-points, are similar.

The $\xi^2$ dependence of the charge transfer is a result of assuming at most a linear DOS variation in the simple
model. In reality, there are more complex
variations of the DOS and its derivatives around $E_F$. 
By imposing a $\xi$ on the calculated bands rigidly, one finds that $\Delta q_{\xi}$ is proportional to $\xi^2$
only below $\sim$ 1 mRy, it is linear around the interval 2-4 mRy, and it saturates at about 0.007 el./La
for $\xi$ larger than about 5 mRy. By comparing the energies for large $\xi$ one can estimate the maximum
$\xi$ for a stable moment:

\begin{equation}
\label{eq:modston} 
N^2 I_s \xi^2+ U_0 0.007  =  N \xi^2    
\end{equation}

With $U_0$=1.5, this gives $\xi$=16 mRy and a moment of $\sim$ 0.8 $\mu_B$ per cell. 
These values are much larger than the values
from the stable spin-polarized solution, about 3 mRy for $\xi$ and 0.1 $\mu_B$ for the moment, located
on La only. The model can reproduce the essential dependencies of the magnetic instability,
and it helps to understand the reasons behind the transition. But for quantitative values one has to
rely on the spin-polarized calculations.

The magnetic state in the supercell is fragile, and in a smaller cell not even stable. The fact that
an insufficient number of k-points can make the moment to increase, is not meant to represent
the real physical situation. It serves to demonstrate that the moment depends on the derivative
of the DOS, and thus to confirm the hypothesis that the charge transfer contributes to the stabilization
of the magnetic state. The magnetism is in this case sensitive to other features in the electronic structure
then what is normal for ferromagnets. A large Stoner factor needs a large DOS, usually a peak, which often
is sensitive to disorder and thermal smearing. But a derivative of the DOS can be large in a fairly
wide energy interval, aside a peak in the DOS. The contribution to magnetism from charge transfer energies
may therefore resist well to thermal effects, and thereby be consistent with the observation of high
Curie temperatures.

 When it comes to explain the properties of the real hexaboride system
from the band results, one has to rely on extrapolations.  
The very low doping ($x$=0.01) for optimal magnetism in real La$_x$Ca$_{(1-x)}$B$_6$ \cite{you},
is much smaller than the doping in the largest supercell ($x$=0.125) studied here. Calculations
for realistically large supercells are not possible at present. 
However, there are several indications that the mechanism favorable to magnetism will remain
or even be stronger at lower doping. 
First,  by increasing the La content
to $x=\frac{1}{2}$ in a double cell LaSrB$_{12}$, one finds a more rigid-band like La-DOS than for
the dilute impurity in the supercell,
and the local La-DOS resembles more the Sr-DOS. Effects of charge transfers will therefore
be less pronounced than for a single La-impurity, and no magnetic state was found in this case. 
Second, the increasing La-La
distance for less doping, will maintain a narrow, impurity-like DOS localized on La. The Fermi-energy
will not cut this band in a rigid-band manner at a lower energy, where the DOS and its derivative is smaller,
even though the number of electrons per average site is decreasing with lower doping. 
Instead, since the additional charge remains localized at the impurity, one expects that the position of $E_F$
remains rather fixed relative to the La-band, with large local values
of $N_{La}$ and $N'_{La}$ even for lower doping. 
 On the other hand, in the large volume far from the La-site one expects
that charges and DOS conditions will be as for the undoped material, 
not knowing about the effects of La in part of the system.
In that case $N_v$ will approach $N$, where in $N$ only the part of the DOS which is affected
by the impurity should enter in eq. \ref{eq:msto}, and the correction to the Stoner factor will remain large.

The appearance of ferromagnetism is therefore a natural consequence of the energy changes associated
with an exchange splitting. These energies can be estimated within a mean-field theory. Favorable
conditions for magnetism are non-constant DOS at $E_F$ to make transfers possible, and a negative
change in total energy as function of the transfers. The non-rigid band like electronic
structure around the dilute La impurity shows up in the distribution of the ferromagnetic moment.
According to these results, it is mainly La and the nearest sites which takes the major part
of the total moment. B-sites remain essentially non-polarized. Alternative theories based on
the excitonic mechanism \cite{ric} need to take into account the non-rigid band effect
of the La-doping. If a rigid-band mechanism did apply, one should expect a uniform
distribution of the moment. Theories of spontaneous ferromagnetism of the dilute electron gas
have also been proposed \cite{ort,you}. However, the very large electron-gas parameter $r_s$ 
necessary for this ($\geq$ 30) is much too large. It is the total (and not only the part due to doping)
electron density that should be considered, and 
for this type of compound $r_s$ is smaller than 1.5 within the atomic sites and about 3 within the empty sites. 

What is shown here is that ferromagnetism occurs in the supercell of LaSr$_7$B$_{48}$. This is despite
the fact that according to the low Stoner factor,  ferromagnetism should not appear.
By doing the calculations with different number of k-points it has been possible to show that the
actual derivative of the local DOS at $E_F$, which is conditional for charge transfers, scales with
the stability of the magnetic moment. This demonstrates that the mechanism for a ferromagnetic
is not exclusively found in the exchange energy.  The necessary condition of a localized, 
non-constant La-DOS near $E_F$,
appears to be enforced as the doping decreases, making the mechanism of charge transfers to persist
at lower doping.  A computational verification of this will need a very large unit cell or a
Green-function method for the limit of dilute doping.




\newpage

\vspace{1.5cm}
\begin{table}
\caption{Paramagnetic DOS values (states/cell Ry, and states/cell Ry$^2$),
 spin-polarized exchange splitting on La $\xi$ (mRy) and magnetic moment on La and total ($\mu_B$),
 as function of the number of k-points. To the impurity site is counted also the contributions
from the 6 closest empty spheres.}
\begin{tabular}{ccccccc} 
  k-points & $N(E_F)$ & $N_{La}$ & $N'_{La}$ & $\xi_{La}$ & $m_{La}$ & $m_{tot}$  \\ 
 4 & 98 & 50 & 2100 & 7  & 0.17   &  0.24 \\
 10 & 52 & 23 & 830 & 4  & 0.10   &  0.15 \\
 20 & 48 & 19 & 730 & 3  & 0.07   &  0.10 \\
\end{tabular}
\label{tab:tm}

\end{table}

\begin{table}
\caption{Differences in valence charge (el. per site) 
between doped and undoped SrB$_6$-supercell for selected sites
calculated from 20 k-points. To the impurity site is counted also the charge
within the 6 closest empty spheres. }
\begin{tabular}{cccc} 
  Impurity-site (La or Sr) & Sr near La & Sr far from La & B   \\ 
 1.19 & 0.005 & 0.01 & -0.005  \\
 
\end{tabular}
\label{tab:ch}

\end{table}

\newpage


\begin{figure}
\caption{Paramagnetic, total and partial La-DOS near $E_F$ for LaSr$_7$B$_{48}$
and total DOS in Sr$_8$B$_{48}$ as obtained from tetrahedron integration from
20 k-points. The zero energy is at $E_F$ for LaSr$_7$B$_{48}$ and
at the energy corresponding to 1 additional electron above $E_F$ for Sr$_8$B$_{48}$.
 \label{fig:dos}}
\end{figure}


\begin{thebibliography}{99}

\bibitem{ott1} H.R. Ott {\it et al}, Z. Phys. B {\bf 102}, 337 (1997);
 H.R. Ott {\it et al}, World Scientific (1998).
 
\bibitem{deg} L Degiorgi {\it et al}, Phys. Rev. Lett {\bf 79}, 5134, (1997).
 
\bibitem{you} D.P. Young {\it et al}, Nature {\bf 397}, 412 (1999);

\bibitem{mass} S. Massidda {\it et al}, Z. Phys. B {\bf 102}, 83, (1997).

\bibitem{lda} W. Kohn and L.J. Sham, Phys. Rev. {\bf 140}, A1133, (1965).

\bibitem{hunt} E.C. Stoner, Rep. Prog. Phys. {\bf 11}, 43 (1948);  K.L. Hunt,
Roc. Roy. Soc. A{\bf 216}, 103, (1953).

\bibitem{jan} O. Gunnarsson, J. Phys. F {\bf 6}, 587 (1976);
 J.F. Janak, Phys. Rev. {\bf B16}, 255, (1977).

\bibitem{c15} T. Jarlborg and A. J. Freeman, Phys. Rev. B {\bf 22}, 2332, (1980).

\bibitem{ric} M.E. Zhitomirsky, T. M. Rice and V.I. Anisimov, (cond-mat 9904330) (1999);
 L. Balents and C.M. Varma, (cond-mat 9906259) (1999);
 V. Barzykin and L.P. Gorkov, (cond-mat 9906401) (1999).

\bibitem{ort} G. Ortiz, M. Harris and P. Ballone, Phys. Rev. Lett {\bf 82}, 5317, (1999).

\bibitem{tj98} T. Jarlborg, Phys. Rev. B {\bf 58}, 9599, (1998).








\end{thebibliography}
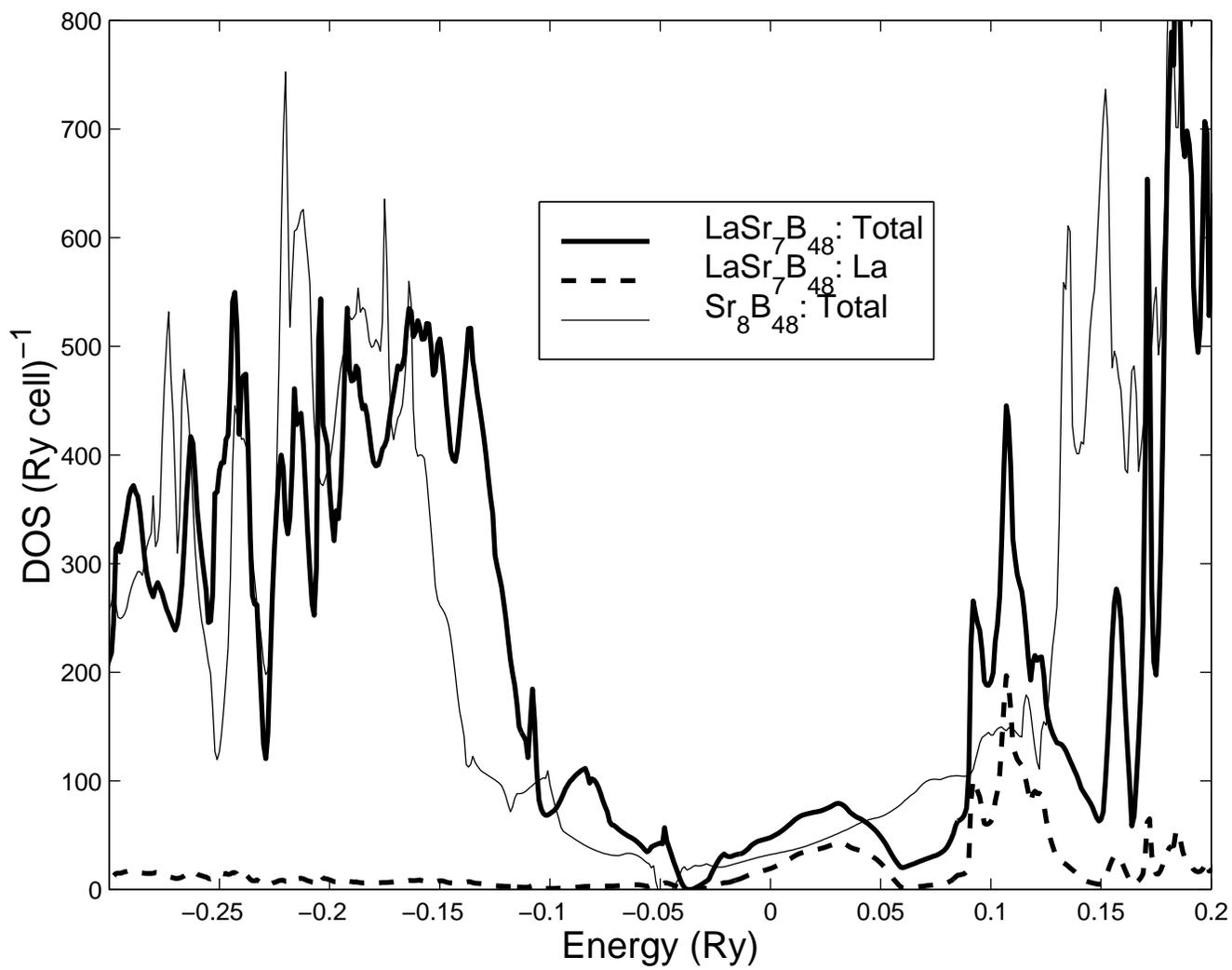
\end{document}